\documentstyle[12pt]{article}

\begin{document}

\begin{center}

{}~\vfill

{\large \bf  Ambiguity of the one-loop calculations  \protect \\
in a nonrenormalizable quantum gravity}

\vfill

{\large M. Yu. Kalmykov}
\footnote { E-mail: $kalmykov@thsun1.jinr.dubna.su$
 \protect \\
Supported in part by ISF grant \# RFL000}

\vspace{2cm}

{\em Bogoliubov Laboratory of Theoretical Physics,
 Joint Institute for Nuclear  Research,
 $141~980$ Dubna $($Moscow Region$)$, Russian Federation}

\vspace{1cm}

and

\vspace{1cm}

{\large P. I. Pronin}\footnote{E-mail: $pronin@theor.phys.msu.su$}

\vspace{1cm}

{\em Department of Theoretical Physics, Physics Faculty\\
 Moscow State University, $117234$, Moscow, Russian  Federation}

\end{center}

\vfill

\begin{abstract}
The ambiguity in the calculations of one-loop counterterms by
the background field method in nonrenormalizable theories of gravity is
discussed. Some examples of such ambiguous calculations are given.  The
non-equivalence of the first and second order formalism in the
quantum gravity is shown.
\end{abstract}

{}~\vfill

PCAS number(s) 04.60.-m, 11.10.Gh

\pagebreak

\section{Introduction}

The construction of a quantum theory of gravity is an unresolved
problem of modern theoretical physics. The Einstein gravity satisfies
to the four classical experimental verifications~\cite{will};
however, this theory is incomplete.  Yet at the classical level
there are problems connected with the presence of the
singularity ~\cite{sin1} - ~\cite{sin3}, the definition of the
energy-momentum tensor of gravitational field, etc.
The main problems are connected with the quantum treatment of the
gravitational interaction. Einstein's gravity is the finite theory
at the one-loop level in the absence of both matter fields and a
cosmological constant~\cite{THV}, but it is  nonrenormalizable
theory at the two-loop order~\cite{GS},~\cite{V}.  The
interaction of the gravity with the matter fields gives rise to
nonrenormalizable theories yet at the one loop level
{}~\cite{DPN1}-~\cite{DPN3}.  Therefore, one needs to modify the theory
or to show that difficulties presently encountered by the theory are
only artifacts of the perturbation theory. For example, one can reject
the perturbation renormalization of the theory as a main criterion of
the true quantum gravity theory. We can consider  the finite criterion:
the Green functions can be divergent but all elements of the S-matrix
must be finite on the mass-shell in each order of perturbation theory.
This criterion is satisfactory in the supergravity theories.

Now there are several ways to modify the Einstein's gravity. The most
interesting directions are following:

\begin{enumerate}
\item
One can introduce terms  quad\-ra\-tic in the curvature tensor in the
action of the theory. This theory is renormalizable but it is not
unitary because the ghosts and tachyons are present in the spectrum of
the theory~\cite{St2} - ~\cite{John}. It is impossible to restore the
unitarity of the theory by loop corrections or adding an interaction
with matter fields (see also ~\cite{Renor}).
\item
One can consider the non-Riemannian geometry. This way is connected
with the possibility for the quantum (and possible classical)
treatment of space-time to involve more than the Riemannian space-time
\cite{FWH1} - \cite{FWH6}.  The most interesting non-Riemannian
space-times are the space-time with torsion and affine-metric
space-time. In these geometries, there are geometrical objects
additional to the metric tensor such as torsion and nonmetricity
tensors defined as independent variables.  In these theories, there are
additional symmetries connected with the local transformation of the
connection fields \cite{MKL2}, \cite{MKL3}. The presence of
additional symmetries in the theory may improve the renormalization
properties of the theory. However, all attempts to construct the
perturbative renormalizable and unitary quantum gravity based on the
non-Riemannian space-time failed.
\item
One can consider the theories with an additional gauge symmetries.
The most promising symmetry is supersymmetry. In supergravity we
must use the finite criterion for obtaining the sensible quantum
gravity. The simplest supergravity with $N=1$ is the first example
of the gravity theory interacting with the matter field which has the
finite elements of the S-matrix on the mass-shell at the one-loop
level.  But at the three-loop level, there are nonvanishing
counterterms violating the finiteness of the theory. The extended
supergravities with $N > 1$ may be finite up to $N$ loop. But at the
present time, there is not a satisfactory supergravity model finite at
all loop levels ~\cite{SUGRA3}, ~\cite{SUGRA4}.  \end{enumerate}

In recent years, hopes of constructing a renormalizable theory of the
quantum gravity have centered on the  superstring ~\cite{GSW}. The
question about the existence of a perturbative renormalizable quantum
gravity in any string model is open.

The new promising non-perturbative treatment of the Einstein gravity
is discussed in ~\cite{A1}.

Modern quantum field theory is based on the principles like
unitarity, renormalizability, the existence of the S-matrix and
perturbation approach.
All suggested models of the quantum gravity based on the
Riemannian or non-Riemannian geometries can be divided into three
classes:
\begin{enumerate}
\item the renormalizable, but non-unitary models
\item unitary, but nonrenormalizable models
\item nonrenormalizable and non-unitary models
\end{enumerate}
Hence, all existing theories of the quantum
gravity are unsatisfactory from the point of view of quantum field
theory. In the gravity, quantum corrections  give rise to very
interesting results like the modification of the Newton law,
disappearance of the classical singularity, corrections to the entropy
of the black hole.  All these results were obtained by means of the
quantum field theory methods.  Since all existing theories of gravity
are unsatisfactory from the point of view of quantum field theory , the
question arises about the validity of the results of loop
calculations. In other words, one needs to investigate the consistency
of the modern powerful tool of quantum field theory and existing
theories of gravity.

In this paper, we will discuss only the validity of the results
of one-loop calculations in the framework of the background field
method in nonrenormalizable theories of the  quantum gravity.
We will concentrate our attention on the DeWitt-Kallosh and
the equivalence theorems, which play the essential role in the modern
methods of the loop calculations in quantum gravity.

The equivalence theorem states, that the $S$-matrix of the
renormalizable theory is independent of the following change of
variables:

\begin{equation}
\varphi^j \rightarrow  '\varphi^j = \varphi^j +
\left( \varphi^2 \right)^j + \left( \varphi^3 \right)^j + \dots
\label{smat}
\end{equation}

In the case of the quantum gravity, this statement is divided into two
parts:

\begin{enumerate}
\item It is well know that there is considerable freedom in what one
considers to be gravitational fields. For example, in the Einstein
gravity we can consider an arbitrary tensor density  $\tilde{g}_{\mu
\nu } = g_{\mu \nu }(-g)^r$ or $\tilde{g}^{\mu \nu } = g^{\mu \nu
}(-g)^s$ as gravitational variables. In accordance with the equivalence
theorem the loop counterterms on the mass-shell must be independent of
the choice of gravitational variables.  \item The loop counterterms on
the mass-shell are independent of the redefinition of quantum fields of
the form $$ h_{\mu \nu } \rightarrow 'h_{\mu \nu }  = h_{\mu \nu }  +
{\it k} \left( h^2 \right)_{\mu \nu } + {\it k}^2 \left( h^3
\right)_{\mu \nu } + \dots $$ This redefinition must influence only the
higher loop results off the mass-shell.  \end{enumerate}

By means of the corresponding choice of gravitational variables or
the corresponding quantum field redefinition, one can considerably
reduce the number and the type of interaction vertices. For example, if
we consider $g_{\mu \nu }$ as a gravitational variable, the number of
three-point interactions in the Einstein gravity is equal to
13~\cite{GS}; if the tensor density $g^{\mu \nu } \sqrt{-g}$ is
selected as a dynamical variable, the number of a three-point
interaction is equal to six~\cite{CLR}; combining both the method
reduces the number of three-point interactions to two~\cite{V}.

The main aim of our investigation is to show that the results of
loop calculations within the background field method in
nonrenormalizable theories of quantum gravity are ambiguous.
As a consequence, we assert that in the nonrenormalizable theories of
the quantum gravity the usual (background) effective action
on and off shell does not give physical information.

We use the following notation:
$$ c = \hbar = 1;~~~~~  {\it k}^2 = 16 \pi G,
{}~~~~~g = -det(g_{\mu \nu }),
{}~~~~~e = \left| det(e^a_{~ \mu}) \right| $$
$$\eta_{\mu \nu} = (+ - - - ),
{}~~~~~\varepsilon  = \frac{4-d}{2},
{}~~~~~\mu , \nu  = 0,1,2,3; ~~~~~ a,b = 0,1,2,3;$$
$$ R^\sigma _{~\lambda  \mu  \nu } = \partial_\mu \Gamma^\sigma
_{~\lambda \nu }  - \partial_\nu \Gamma^\sigma _{\lambda \mu } +
\Gamma^\sigma_{~\alpha \mu } \Gamma^\alpha_{~\lambda  \nu } -
\Gamma^\sigma_{~\alpha  \nu }  \Gamma^\alpha_{\lambda \mu },~~~~~
R_{\mu \nu } = R^\sigma_{~\mu \sigma \nu },~~~~~
  R =  R_{~\mu \nu } g^{\mu \nu } $$
where $\Gamma^\sigma_{~\mu \nu } $ is the Riemannian connection defined as

\begin{equation}
\Gamma^\sigma_{~\mu\nu}  = \frac{1}{2} g^{\sigma \lambda} \left(
- \partial_\lambda g_{\mu \nu} + \partial_\mu  g_{\nu \lambda} +
\partial_\nu g_{\lambda \mu} \right)
\label{Rieman}
\end{equation}

Objects marked by the tilde $\tilde{}$ are constructed by means of the
Riemann-Cartan connection $\tilde \Gamma^\sigma_{~\mu \nu}$.
Other objects are the Riemannian objects.

\section{Background field method}

The background field method ~\cite{BDW-67},~\cite{BDW} was suggested
to obtain  covariant results of the loop calculations. In the
background field method, all dynamical fields $\varphi^j $ are expanded
with respect to background values, according to $$ \varphi^j =
\varphi^j_b  + \phi^j_q $$ and only the quantum fields $\phi^j_q$ are
integrated over in the path integral.  The background fields
$\varphi^j_b$ are effectively external sources.  For the
one-particle irreducible diagrams there is a difference between the
normal field theory and the background field method insofar as the
gauge-fixing term may introduce additional vertices.  B.DeWitt has
proved that these additional vertices do not influence  the
$S$-matrix and the $S$-matrix in the formalism of the background field
method is equivalent to the conventional $S$-matrix
{}~\cite{BDW-67},~\cite{BDW}.  This proof has later been extended
in a lot of papers ~\cite{TH1} - ~\cite{vor3}. The physical quantities
are gauge and parametrization independent elements of the S-matrix on
the mass shell. The choice of  external lines on the mass-shell in
the background field formalism corresponds to using the classical
equations of motion for the background fields. Hence, the counterterms
on the mass-shell calculated by the background field method must be
independent of the gauge-fixing parameters and the reparametrization of
quantum fields. These statements are called the DeWitt-Kallosh
theorem ~\cite{BDW-67},~\cite{BDW},~\cite{Kal} and equivalence theorem
{}~\cite{equiv1}-~\cite{equiv4}, respectively. However, for
nonrenormalizable theories the proofs of the DeWitt-Kallosh theorem and
equivalence theorem are formal.

In the next chapter, we restrict ourselves only to one-loop
calculations. Let us give the short notes about the one-loop
calculations within the background field formalism.

In the gauge theories, the
renormalization procedure may violate the gauge invariance at the
quantum level, thus destroying the renormalizability of the theory.
Therefore, one is bound to apply an invariant renormalization. This can
be achieved by applying an invariant regularization and using the
minimal subtraction scheme ~\cite{min},~\cite{Sp}. It has been proved
that the dimensional regularization ~\cite{dim1} - ~\cite{dim4} is an
invariant regularization preserving all the symmetries of the classical
action that do not depend explicitly on the space-time dimension
\cite{Sp},~\cite{action},~\cite{vlad}.  It has been shown ~\cite{AKK}
that in general renormalizable and nonrenormalizable theories the
background field formalism requires using an invariant
renormalization procedure to  obtain valid results. A
noninvariant regularization or renormalization may break an implicit
correlation between different diagrams, which is essential as one
formally expands the action in the background and quantum fields. We
will use the invariant regularization (dimensional renormalization and
minimal subtraction scheme) in our calculations.

Let us consider the gauge theory with the classical action
$S(\phi^j )$ where \{$\phi^j$\} are the dynamical variables. In
accordance with the background field method, all dynamical fields
$\varphi^j $ are rewritten as a sum of the background and quantum
fields:

\begin{equation}
\phi^j = \phi_b^j + \phi_{qu}^j
\end{equation}
and the fields $\phi_b^j$ satisfy the classical equations of motion

\begin{equation}
\frac{\delta S(\phi_a^i)}{\delta \phi_b^j}  = 0
\end{equation}

One expands the action $S(\varphi_b^j + \varphi_{qu}^j )$ in powers of
the quantum field and picks out the terms quadratic in the quantum
fields we obtain

\begin{equation}
S_{eff}  = \frac{1}{2}  \phi^i_{qu}
\frac{\delta^2 S(\phi_b)}{\delta \phi^i_b \delta \phi^j_b} \phi^j_{qu}
\end{equation}

This is an effective action for  calculating the one -loop corrections.
Due to the presence of the gauge invariance, one needs to introduce the
gauge fixing term

\begin{equation}
f^a = P^a_j(\phi_b ) \phi^j_{qu}
\end{equation}

\begin{equation}
S_{gf} = \frac{1}{2\alpha } f^af_a
\end{equation}
where $\alpha$ is an arbitrary constant and $P^a_j(\phi_b )$ is the most
general gauge in the background field method defined by the following
conditions:

\begin{itemize}
\item Lorentz covariance
\item linear in the quantum field
\item the number of derivatives with respect to  the quantum fields is
smaller than or equal to one
\end{itemize}

When using the invariant renormalization the one-loop correction to the
usual effective action is

\begin{equation}
\Gamma^{(1)}  = \frac{i}{2} \left( \ln {\it det} \triangle_{ab} -
2 \ln {\it det} \triangle_{FP} \right)
\label{one-loop}
\end{equation}
where

$\triangle_{FP}$ is the Faddeev-Popov ghost operator, defined in the
standard way and

\begin{equation}
\triangle_{ij}  = \frac{\delta^2 S(\phi)}{\delta \phi^i \delta
\phi^j} + P^a_{i}(\phi ) P_{aj}(\phi )
\label{def}
\end{equation}

The divergence part of the one-loop effective action  obtained by
means of the heat kernel method is

\begin{equation}
\Gamma^{(1)}_\infty  = -\frac{1}{32 \pi^2 \varepsilon }
\int d^4 x \sqrt{-g} \bigl( B_4(\triangle_{ij}) - 2 B_4(\triangle_{FP})
\bigr)
\label{div}
\end{equation}
\noindent where $B_4$  is the second coefficient of the
spectral expansion of the corresponding differential operator
{}~\cite{Gil} - ~\cite{Bar}. For the operator

\begin{equation}
\triangle_{ij} = - \left(  \nabla^2 {\bf 1}_{ij} + 2 S^\sigma_{~ij}
\nabla_\sigma + X_{ij} \right)
\label{oper}
\end{equation}

$B_4$ is equal to

\begin{equation}
B_4(\triangle) = {\it Tr} \Biggl( \frac{1}{180}
\left( R^2_{\mu \nu \sigma \lambda } - R^2_{\mu \nu } \right)
+ \frac{1}{2} \left( Z + \frac{R}{6} \right) + \frac{1}{12}
Y_{\mu \nu } Y^{\mu \nu }  +
\frac{1}{6} \Box \left( \frac{R}{5} + Z \right) \Biggr)
\label{B}
\end{equation}

where

\begin{eqnarray}
Z & = & X - \nabla_\lambda S^\lambda - S_\lambda S^\lambda  \nonumber \\
Y_{\mu \nu }  & = & \nabla_\mu S_\nu - \nabla_\nu S_\mu  +
S_\mu S_\nu  - S_\nu S_\mu  + [\nabla_\mu , \nabla_\nu ]{\bf 1}
\end{eqnarray}

\section{Examples of ambiguity of the one-loop calculations}

In this section, we remind some previous results on the
ambiguity in one-loop calculations in nonrenormalizable
theories of gravity (Examples $1$ and $2$)

\subsection{Example 1}

Let us consider the matter field in the external gravitational
background \cite{Duff}. Consider the interaction of the gravity based
on the Riemannian space-time with a real scalar field $\phi$ described
by the action

\begin{equation}
S_1(g, \phi)  = \int d^4 x~ g^{\mu \nu} \sqrt{g}  \left(
\frac{1}{2} \partial_{\mu} \phi \partial_{\nu} \phi
- \frac{1}{{\it k}^2} R_{\mu\nu} (g)
+ \frac{1}{12} \phi^2 R_{\mu\nu} (g) \right)
\label{D1}
\end{equation}

Now we make the change of the variables

\begin{equation}
g_{\mu \nu}  = G_{\mu \nu}  cosh^2 \left( {\it k} \psi \right)
\label{D2}
\end{equation}

\begin{equation}
\phi = {\it k}^{-1} tanh \left( {\it k} \psi \right)
\label{D3}
\end{equation}

The new action is given by

\begin{equation}
S_2(G, \psi)  = \int d^4 x~ G^{\mu \nu} \sqrt{G}  \left(
\frac{1}{2} \partial_{\mu} \psi \partial_{\nu} \psi
- \frac{1}{{\it k}^2} R_{\mu\nu} (G) \right)
\label{D4}
\end{equation}

In accordance with the equivalence theorem, the S-matrix corresponding
to the action $S_1$ coincides with the S-matrix corresponding to the
action $S_2$. The DeWitt-Kallosh theorem asserts that the one-loop
counterterms on the mass-shell calculated by the background field
method must be gauge and parametrization invariant. As consequence, the
one-loop counterterms of the theory described by the action $S_1$ must
coincide with the one-loop counterterms of the theory described by the
action $S_2$. As has been shown by M. J. Duff \cite{Duff} when only the
scalar field is quantized, which corresponds to quantum field theory
in the external curved space-time, the one-loop counterterms are

\begin{equation}
\triangle ç_{1 \infty}  = \frac{1}{1920 \pi^2 \varepsilon}
\int d^4 x~ \sqrt{g}~ C_{\alpha \beta \mu \nu}(g)
C^{\alpha \beta \mu \nu}(g)
\label{D5}
\end{equation}
and

\begin{equation}
\triangle ç_{2 \infty}  = \frac{1}{1920 \pi^2 \varepsilon}
\int d^4 x~ \sqrt{G} \left( C_{\alpha \beta \mu \nu}(G)
C^{\alpha \beta \mu \nu}(G)  + \frac{5}{2} R^2(G) \right)
\label{D6}
\end{equation}
for the actions $S_1$ and $S_2$, respectively. Here
$C_{\alpha \beta \mu \nu}$ is the Weyl tensor.

Since
\begin{equation}
C_{\alpha \beta \mu \nu}(g) C^{\alpha \beta \mu \nu}(g) \sqrt{g}  =
C_{\alpha \beta \mu \nu}(G) C^{\alpha \beta \mu \nu}(G) \sqrt{G}
\label{D7}
\end{equation}
we see that

\begin{equation}
\triangle ç_{1 \infty} \neq \triangle ç_{2 \infty}
\label{D8}
\end{equation}

Hence, the equivalence theorem is violated.

\subsection{Example 2}

Consider the interaction of the Einstein gravity with the scalar field
$\phi$  described by the action

\begin{equation}
S (g, \phi)  = \int d^4 x~ g^{\mu \nu} \sqrt{g}  \left(
\frac{1}{2} \partial_{\mu} \phi \partial_{\nu} \phi
- \frac{1}{{\it k}^2} R_{\mu\nu} (g) \right)
\label{I1}
\end{equation}

For the calculation of the one-loop counterterms within the
background field me\-thod, we use the following gauge \cite{Ichi1}:

\begin{equation}
L_{gf}  = \frac{1}{2 \beta_1} C_\mu C^\mu
\label{I2}
\end{equation}

\begin{equation}
C_\mu = \nabla_\nu h_\mu^\nu - \frac{1}{2} \alpha_1 \nabla_\mu h
- \alpha_2 {\it k} \varphi \partial_\mu \phi
- \alpha_3 {\it k} \phi \partial_\mu \varphi
\label{I3}
\end{equation}
where $h_{\mu \nu}$ and $\varphi$ are the quantum metric and
scalar fields, respectively, and
$$
\begin{array}{ccc}
\alpha_1  = 1 - 2\alpha, &
\beta_1  =  1 - 2 \beta, &
\alpha_2 = 1 + \xi  \\
\left| \alpha \right| \ll 1, &
\left| \beta \right| \ll 1,  &
\left| \alpha_3 \right| \ll 1
\end{array}
$$
and $\xi$ is arbitrary.

The one-loop counterterms on the mass-shell in the topological trivial
space-time are

\begin{eqnarray}
\triangle ç_{\infty}  & = & \frac{1}{8 \pi^2 \varepsilon} \frac{1}{5760}
\int d^4 x~ \sqrt{g} \left(\partial_\mu \phi \partial_\mu \phi \right)^2
\Bigl(-3654 + 60 \alpha (-3 \xi^4 + 24 \xi^3 - 35 \xi^2 + 11 \xi)
\nonumber \\
&& + 720 \alpha_3 (-\xi + 2)
+ 5 \beta (90 \xi^4 - 273 \xi^2 + 324 \xi + 116)
+ O(\alpha, \alpha_3, \beta)^2 \Bigr)
\label{I4}
\end{eqnarray}

Hence the DeWitt-Kallosh theorem is violated.

\section{One-loop counterterms in the first-order gravity with the
Gilbert-Einstein action}

Let us consider the Riemann-Cartan space-time. Let us a brifly
describe the mathematical tool of the space-time with the
torsion. In the Riemann-Cartan space-time there are two
equivalent approaches for describing the geometry of the space-time with
torsion.

The first approach, so-called the Poincar\`{e} gauge approach,
considers the vierbein field $e^a_{~\mu}$ and local Lorentz connection
$\tilde w^a_{~b \mu}$ as independent dynamical variables. In this
approach, there are two gauge field strengths. One is the translational
gauge field strength defined by

\begin{equation}
Q^a_{~\mu \nu}(e, \tilde w)  \equiv - \frac{1}{2}
\left( \partial_\mu e^a_{~\nu}  - \partial_\nu e^a_{~\mu}
+ \tilde w^a_{~b \mu} e^b_{~\nu}  - \tilde w^a_{~b \nu} e^b_{~\mu}
\right)
\label{tors}
\end{equation}

This tensor is the strength tensor of the vierbein $e^a_{~\mu}$. The
other is the Lorentz gauge field strength defined by the following
relation:

\begin{equation}
\tilde R^a_{~b \mu \nu}(\tilde w)  \equiv
\partial_\mu \tilde w^a_{~b \nu} - \partial_\nu \tilde w^a_{~b \mu}
+ \tilde w^a_{~c \mu} \tilde w^c_{~b \nu}
- \tilde w^a_{~c \nu} \tilde w^c_{~b \mu}
\label{curvat}
\end{equation}

The second approach to description of the Riemannian-Cartan space-time,
so-called geometrical approach, considers the metric tensor $g_{\mu
\nu}$ and linear affine connection $\tilde \Gamma^\sigma_{~\mu \nu}$ as
independent dynamical variables. These variables satisfy the metric
condition

\begin{equation}
\tilde \nabla_\sigma g_{\mu \nu} \equiv \partial_\sigma g_{\mu \nu}
- \tilde \Gamma^\alpha_{~\mu \sigma} g_{\alpha \nu}
- \tilde \Gamma^\alpha_{~\nu \sigma} g_{\alpha \mu}  = 0
\label{metric}
\end{equation}

By means of these variables we can define two geometrical objects, the
curvature and the torsion tensors, characterizing the Riemann-Cartan
space-time. The torsion and curvature tensors are defined by the
following expressions:

\begin{equation}
Q^\sigma_{~\mu \nu}(\tilde \Gamma )  \equiv \frac{1}{2} \left(
\tilde \Gamma^\sigma_{~\mu \nu} - \tilde \Gamma^\sigma_{~ \nu \mu}
\right)
\label{tors2}
\end{equation}

\begin{equation}
\tilde R^\sigma_{~\lambda \mu \nu} (\tilde \Gamma) \equiv
  \partial_\mu \tilde \Gamma^\sigma_{~\lambda \nu}
- \partial_\nu \tilde \Gamma^\sigma_{~\lambda \mu}
+ \tilde \Gamma^\sigma_{~\alpha \mu} \tilde \Gamma^\alpha_{~\lambda \nu}
- \tilde \Gamma^\sigma_{~\alpha \nu} \tilde \Gamma^\alpha_{~\lambda \mu}
\label{curvat2}
\end{equation}

The Poincar\`{e} gauge approach and geometrical approach are related
with each other by the following constraint equations:

\begin{equation}
g_{\mu \nu}  = e^a_{~\mu} e^b_{~\nu} \eta_{a b}
\label{const1}
\end{equation}

\begin{equation}
\tilde \nabla_\sigma e^a_{~\mu} \equiv \partial_\sigma e^a_{~\mu}
+ \tilde w^a_{~b \sigma} e^b_{~\mu}
- \tilde \Gamma^\lambda_{~\mu \sigma} e^a_{~\lambda}  = 0
\label{const2}
\end{equation}

Due to equations (\ref{const1}) and (\ref{const2}) the connection
between the Poincar\`{e} gauge approach and the geometrical approach
becomes very clear. The translational gauge field strength $Q^a_{\mu
\nu}(e,\tilde w)$ is the torsion field

\begin{equation}
Q^a_{~\mu \nu}(e, \tilde w)  =
e^a_{~\sigma} Q^\sigma_{~\mu \nu}(\tilde \Gamma)
\label{conn1}
\end{equation}
and the Lorentz gauge field strength $\tilde R^a_{~b \mu \nu}(\tilde w)$
is the curvature tensor

\begin{equation}
\tilde R^a_{~b \mu \nu}(\tilde w)  = e^a_{~\sigma} e_b^{~\lambda}
\tilde R^\sigma_{~\lambda \mu \nu}(\tilde \Gamma)
\label{conn2}
\end{equation}

Greek and Latin indices is converted into Latin or Greek indices with
the help of the vierbein field, for example

\begin{equation}
K^a_{~b \nu}  = e^a_{~\sigma} e_b^{~\mu} K^\sigma_{~\mu \nu}
\label{conn3}
\end{equation}

Having solved equation (\ref{metric}) and using the constrain equations
(\ref{const1}) and (\ref{const2})
we obtain the following decompositions of the linear affine connection
$\tilde \Gamma^\sigma_{~\mu \nu } $ and local Lorentz connection $\tilde
w^a_{~b \mu }$:

\begin{equation}
\tilde \Gamma^\sigma_{~\mu \nu }  = \Gamma^\sigma_{~\mu \nu }
+ K^\sigma_{~\mu \nu }
\label{decom1}
\end{equation}

\begin{equation}
\tilde w^a_{~b \mu }  = w^a_{~b \mu } + K^a_{~b \mu }
\label{decom2}
\end{equation}
where

$\Gamma^\sigma_{~\mu \nu } $  is the Riemannian connection
defined in (\ref{Rieman}),

\begin{equation}
K^\sigma_{~\mu \nu } \equiv Q^\sigma_{~\mu \nu }
+ Q_{\mu \nu }^{~~\sigma } + Q_{\nu \mu }^{~~\sigma }
\label{torcontor}
\end{equation}

$w^a_{~b \mu }$ is the Ricci rotation coefficients given by
first derivatives of the vierbein fields

\begin{equation}
w_{abm}  = C_{abm} + C_{bma} + C_{mba}
\label{Ricci}
\end{equation}
where

\begin{equation}
C_{abm} = \frac{1}{2} e_b^{~\mu } e_m^{~\nu } \left(
\partial_\mu e_{a \nu } - \partial_\nu e_{a \mu } \right)
\label{nongolon}
\end{equation}
The tensor $K^a_{~b \mu }$ is a contorsion tensor defined in
(\ref{conn3})

Let us consider the following action:

\begin{equation}
S_1 =  - \frac{1}{{\it k}^2} \int d^4 x~ e
\left( \tilde R(e, \tilde w) - 2 \Lambda \right)
\label{action}
\end{equation}

Using the decomposition of the Lorentz connection
$\tilde w^a_{~b \mu }$ into its irreducible parts (\ref{decom2}), it is
possible to rewrite the action (\ref{action}) in the following form:

\begin{equation}
S_2 =  - \frac{1}{{\it k}^2} \int d^4 x~ e
\left( R(e) - 2 \Lambda - 4 \nabla_\sigma Q^\sigma - 4 Q^\sigma
Q_\sigma  + Q^{\sigma \mu \nu } Q_{\sigma \mu \nu }
+ 2 Q^{\sigma \mu \nu } Q_{\nu \mu \sigma } \right)
\label{action2}
\end{equation}

The third term in expression (\ref{action2}) is the full derivatives

\begin{equation}
\int d^4 x~\sqrt{g}~ \nabla_\mu Q^\mu  =
\int d^4 x~\partial_\mu \left(\sqrt{g}  Q^\mu \right)
\label{deriv}
\end{equation}
and in space-time without boundaries we can neglect this term.

In the Poincar\`{e} gauge approach describing the Riemann-Cartan
space-time there are two sets of dynamical variables:
$\left( e^a_{~\mu} , \tilde w^a_{~b \mu } \right)$ and
$\left( e^a_{~\mu }, Q^a_{\mu \nu } \right)$ (we can consider the
contorsion tensor $K^a_{~\mu \nu }$ instead of the torsion tensor
$Q^a_{~\mu \nu }$). The equations of motion of the theory described by
the action (\ref{action}) or classical equivalent action (\ref{action2})
are independent of the choice of the dynamical variables and have the
following form:

\begin{equation}
R_{\mu \nu }(e)  = \Lambda g_{\mu \nu }
\label{mass-shell1}
\end{equation}

\begin{equation}
Q^a_{~\mu \nu }  = 0
\label{mass-shell2}
\end{equation}

 From equation (\ref{mass-shell2}) we see that the torsion field
$Q^a_{~\mu \nu }$ is a nonpropagating auxiliary field that can be
excluded from the Lagrangian by means of the equation of motion. In
other words, the equation of motion of the torsion tensor $Q^a_{~\mu
\nu }$ is the second-class constraint. So the theories described by the
actions (\ref{action}) and (\ref{action2}) are equivalent to Einstein's
theory with the cosmological constant at the tree level in the absence
of the matter fields. From general consideration of
the theories with constrains \cite{Gitman} one knows that in the
renormalizable theory with the second-class constraints the loop
calculations can be done by the two equivalent methods:

\begin{enumerate}
\item One excludes auxiliary fields from the Lagrangian by means of the
equations of motion (these equations are in general the second-class
constraints) at the classical level and quantizes the obtained theory.
\item One considers auxiliary fields as independent dynamical
variables and quantizes the theory with the existing sets of
fields.
\end{enumerate}

These two methods of calculations give rise to the identical
results of loop calculations in the renormalizable theories. In
particular, the equivalence of quantum theory in the first
and second-order formalism is based on these two
equivalent quantization methods.

We consider two sets of the in\-de\-pen\-dent dy\-na\-mi\-cal
va\-ri\-ab\-les $\left( e^a_{~\mu }, Q^a_{~\mu \nu } \right)$ and
$(e^a_{~\mu }, \tilde w^a_{~b \mu })$. The corresponding actions called
$S_2$ and $S_1$, respectively, are given in (\ref{action2}) and
(\ref{action}).

Let us consider the action $S_2$. Using the first method of
quantization we obtain that after excluding the torsion fields
$Q^a_{~\mu \nu }$ by means of the equation of motion (\ref{mass-shell2})
the action $S_2$ reduces to the ordinary action of the Einstein gravity
with the cosmological constant in the vierbein formalism

\begin{equation}
S_2 \rightarrow  S_2^{mod} =
 - \frac{1}{{\it k}^2} \int d^4 x~ e \left(  R(e) - 2 \Lambda \right)
\label{action2m}
\end{equation}

The vierbein fields have sixteen components: in addition to their
ten (metric) symmetrical components they have six antisymmetrical components,
expressing the freedom of homogeneous transformations of the local Lorentz
frames, which introduce additional dynamical content, especially at
the quantum level. The theory (\ref{action2m}) has two kinds of gauge
invariance: the usual coordinate freedom and the local Lorentz rotations.
Both gauges must be fixed in the covariant quantization scheme by adding
gauge-breaking terms. In the special gauge fixing the local
Lorentz invariance, the contribution of the antisymmetrical vierbein
components and their ghosts disappear from the quantized theory. In
this gauge, the vierbein and metric formulations are equivalent at the
quantum domain in the absence of the spinor fields ~\cite{Spinor}.
Hence, the results of the one-loop calculations in the metric and
vierbein formulations coincide. Using the results obtained in paper
{}~\cite{ChD1} it is possible to write the one-loop counterterms of
the action (\ref{action2m}) in the vierbein formalism.

\begin{equation}
\triangle ç_{\infty} = - \frac{1}{32 \pi^2 \varepsilon }
 \int d^4 x~ e \left(
\frac{53}{45}
R_{\mu \nu \sigma \lambda }(e) R^{\mu \nu \sigma \lambda }(e)
- \frac{58}{5} \Lambda^2 \right)
\label{result}
\end{equation}

Now we consider the second method of quantization. Rewriting all
the dynamical variables as a sum of classical and quantum fields

\begin{equation}
e^a_{~\mu }  = e^a_{~\mu } + {\it k} \lambda^a_{~\mu }
\label{exp1}
\end{equation}

\begin{equation}
Q^a_{~\mu \nu }  = Q^a_{~\mu \nu }  + {\it k} q^a_{~\mu \nu }
\label{exp2}
\end{equation}
and expanding the action $S_2$ in powers of the quantum field up to
terms quadratic in the quantum fields, we obtain the effective
Lagrangian for the calculation of the one-loop counterterms

\begin{equation}
L_{eff}  =  L_{eff}^{GR} - \frac{1}{2}
q^a_{~\mu \nu } H_{a~~b}^{~\mu \nu ~\alpha \beta } q^b_{~\alpha \beta }
 - \frac{1}{2}
\lambda^a_{~\mu } X_{a~b}^{~\mu ~\nu } \lambda^b_{~\nu }
- \lambda^b_{~\sigma  } Z_{b~a}^{~\sigma~\mu \nu  } q^a_{~\mu \nu }
\label{eff1}
\end{equation}
where
$L_{eff}^{GR}$ is the effective Lagrangian of the Einstein gravity with
the cosmological constant quadratic in the quantum field
$\lambda^a_{~\mu } $,
$X_{a~b}^{~\mu ~\nu }$ is proportional to $Q^2$,
$Z_{b~a}^{~\sigma~\mu \nu  }$ is proportional to $Q$ and
$H_{a~~b}^{~\mu \nu ~\alpha \beta }$
consists of the metric tensors and Kronecker's symbols

\begin{eqnarray}
H_{\sigma ~~\rho  }^{~\mu \nu ~\alpha \beta } & = &
  \left(
  \delta^{\beta }_{\sigma } \delta^{\nu }_{\rho } g^{\alpha \mu }
- \delta^{\alpha }_{\sigma } \delta^{\nu }_{\rho } g^{\beta \mu }
- \delta^{\beta }_{\sigma } \delta^{\mu }_{\rho } g^{\alpha \nu }
+ \delta^{\alpha }_{\sigma } \delta^{\mu }_{\rho } g^{\beta \nu }
\right)
+  g_{\sigma \rho }
\left( g^{\alpha \mu } g^{\beta \nu }
- g^{\alpha \nu } g^{\beta \mu } \right)
\nonumber \\
& - & 2 \left(
  \delta^{\beta }_{\rho } \delta^{\nu }_{\sigma } g^{\alpha \mu }
- \delta^{\alpha }_{\rho } \delta^{\nu }_{\sigma } g^{\beta \mu }
- \delta^{\beta }_{\rho } \delta^{\mu }_{\sigma } g^{\alpha \nu }
+ \delta^{\alpha }_{\rho } \delta^{\mu }_{\sigma } g^{\beta \nu }
\right)
\label{H}
\end{eqnarray}

To get the diagonal form of the effective Lagrangian, we are to replace
the dynamical variables in the following way:

\begin{equation}
q^\sigma_{~\mu \nu } \rightarrow  \bar{q}^\sigma_{~\mu \nu } =
q^\sigma_{~\mu \nu } -
H^{-1 \sigma ~~\rho  }_{~~~~\mu \nu ~\alpha \beta }
Z_{b~ \rho  }^{~\tau  ~\alpha \beta   }
\lambda^b_{~\tau }
\label{replacement}
\end{equation}
where $H^{-1 \sigma ~~\rho  }_{~~~~\mu \nu ~\alpha \beta }$
satisfies three conditions

\begin{equation}
H^{-1 \sigma ~~\rho  }_{~~~~\mu \nu ~\alpha \beta } =
H^{-1 \rho  ~~~\sigma }_{~~~~\alpha \beta  ~\mu \nu  }
\label{cond1}
\end{equation}

\begin{equation}
  H^{-1 \sigma ~~\rho  }_{~~~~\mu \nu ~\alpha \beta } =
- H^{-1 \sigma ~~\rho  }_{~~~~\nu \mu ~\alpha \beta } =
- H^{-1 \sigma ~~\rho  }_{~~~~\mu \nu ~\beta \alpha }
\label{cond2}
\end{equation}

\begin{equation}
H^{-1 \sigma ~~\rho  }_{~~~~\mu \nu ~\alpha \beta }
H_{ \rho~~ \omega  }^{~\alpha \beta ~\eta \kappa } =
\frac{1}{2} \delta^\sigma_\omega
\left( \delta_\mu^\eta \delta_\nu^\kappa
- \delta_\mu^\kappa \delta_\nu^\eta \right)
\label{cond3}
\end{equation}

It is known that $H^{-1 \sigma ~~\rho  }_{~~~~\mu \nu ~\alpha \beta }$
satisfying the conditions (\ref{cond1})-(\ref{cond3}) exist
{}~\cite{MKL1}. In the extended theory of gravity additional
symmetries connected with the local transformation of the connection
field may be present in the theory. In this sort of a theory the
expression like $H^{-1 \sigma ~~\rho  }_{~~~~\mu \nu ~\alpha \beta
}$ does not exist ~\cite{MKL2}, ~\cite{MKL3}.

The replacement (\ref{replacement}) does not change the functional
measure

\begin{equation}
det \left|
\frac{\partial (\lambda^a_{~\kappa},  \bar{q}^\sigma_{~\mu \nu })}
{\partial (\lambda^b_{~\tau}, q^\rho_{~\alpha \beta })} \right| = 1
\label{measure}
\end{equation}

Since we are interested only in the results on the mass-shell, it is
possible to consider the effective Lagrangian (\ref{eff1}) and the
replacement of the variables (\ref{replacement}) only on the mass-shell
(\ref{mass-shell2}). Taking into account  the on-shell identities

\begin{equation}
Z_{b~ \rho  }^{~\tau  ~\alpha \beta   } =
X_{a~b}^{~\mu ~\nu } = 0
\end{equation}
we obtain, on the mass-shell, the diagonal effective Lagrangian. The
one-loop counterterms are the sum of the contributions of the quantum
fields $\lambda^a_{~\mu } $ and $q^\sigma_{~\mu \nu }$. On the
mass-shell, the one-loop contribution of the torsion fields to the
effective action is proportional to
$\delta^4(0)det \left(
H_{ \sigma ~~\rho  }^{~\mu \nu ~\alpha \beta } \right)$.
In the dimensional regularization,  $[\delta^4(0)]_R = 0$ and the
one-loop contribution of the torsion fields to the one-loop
counterterms is equal to zero. Hence, the one-loop counterterms depend
only on the contribution of the quantum fields $\lambda^a_{~\mu } $ and
coincide with the standard result (\ref{result}). This result is the
consequence of the equivalence of the two above-mentioned methods of
calculation.

Now we consider the  fields $e^a_{~\mu }$ and $\tilde w^a_{~b \mu }$
as independent dynamical variables. It has been shown
{}~\cite{csecond1} -~\cite{csecond2} that after solving the second-class
constrains that exist in the theory described by the action
(\ref{action}), it is possible to express all $\tilde \omega^a_{~b
\mu } $  as functions of $e^a_\mu $ and exclude $\tilde \omega^a_{~b
\mu }$ as physical degrees of freedom from the theory. Then, the
Lagrangian can be written as

\begin{equation}
S_1 \rightarrow  S_1^{mod} =
 - \frac{1}{{\it k}^2} \int d^4 x~ e \left(  R(e) - 2 \Lambda \right)
\label{actionm}
\end{equation}

The results of the one-loop calculations on the mass-shell are given by
the expression (\ref{result}).

Now we consider the second method of calculation. In accordance with
the background field method, all dynamical variables are rewritten in
the following form:

\begin{eqnarray}
e^a_{~\mu} & = & e^a_{~\mu} + {\it k} \lambda^a_{~\mu}
\nonumber \\
\tilde w^a_{~b \mu} & = & \tilde w^a_{~b \mu} + {\it k} \gamma^a_{~b
\mu}
\label{ex1}
\end{eqnarray}
where $\lambda^a_{~\mu }$ and $\gamma^a_{~b \mu }$ are the quantum
fields and $e^a_{~\mu }$ and $\tilde w^a_{~b \mu }$ are the classical
fields satisfying the equations of motion (\ref{mass-shell1}) and
(\ref{mass-shell2}).

The effective Lagrangian for the calculation of one-loop
counterterms is

\begin{equation}
L_{eff} = - \Biggl( \frac{1}{2} \gamma^a_{~b \mu}
F _{a~~m}^{~b \mu~ n \nu} \gamma^m_{~n \nu}  +
\frac{1}{2} \lambda^b_{~\nu} D_{b~a}^{~\nu~\mu} \lambda^a_{~\mu }
+ \lambda^m_{~n} \biggl( G^{~~n j ~b c}_{m ~~a} \nabla_ j +
T^{~n~b c}_{m~ a} \biggr)
\gamma^a_{~bc} \Biggr) e
\label{eff2}
\end{equation}
where

\begin{equation}
G^{~~nj~bc}_{m~~a} = \delta^n_m \delta^j_a g^{bc} - \delta^n_m
\delta^b_a g^{jc} + \delta^j_m \delta^c_a g^{bn}- \delta^j_m \delta^n_a
g^{bc} + \delta^c_m \delta^n_a g^{jb} - \delta^c_m \delta^j_a g^{bn}
\end{equation}

\begin{equation}
D_{m~e}^{~~n~f} = \biggl( R - 2\Lambda \biggr) \biggl( \delta^n_m
\delta^f_e - \delta^f_m \delta^n_e \biggr)  + 2R^{fn}_{~~me}
 + \frac{1}{2} \biggl(R_m^{~f} \delta^n_e - R^n_{~m} \delta^f_e +
 R_e^{~n} \delta^f_m - R^f_{~e} \delta^n_m  \biggr)
\end{equation}

\begin{eqnarray}
T^{~n~bc}_{m~a} & = & \Biggl(
2 \delta^n_m Q_a g^{bc} - 2 \delta^n_m Q^b \delta^c_a
- \delta^n_m K^{bc}_{~~a} + \delta^n_m K_a^{~cb}
-   g^{bc} K^n_{~am} + \delta^c_a K^{nb}_{~~m}
+ \delta^n_a K^{bc}_{~~m}
\nonumber \\
& - & g^{bn} K^{~c}_{a~m} + \delta^c_m K^{n~b}_{~a}
-  4 \delta^c_m g^{bn} Q_a + 4 \delta^c_m \delta^n_a Q^b
- 2  \delta^c_m K^{nb}_{~~a}
\end{eqnarray}

\begin{eqnarray}
F_{\alpha ~~\mu }^{~\beta \lambda ~\nu \sigma } & = &
\frac{1}{4} \biggl(
   g^{\beta \lambda } \delta^\nu_\alpha  \delta^\sigma_\mu
 - g^{\beta \nu } \delta^\lambda_\alpha  \delta^\sigma_\mu
 - g^{\beta \lambda } g^{\sigma \nu } g_{\alpha \mu }
 + g^{\sigma \nu } \delta^\lambda_\alpha  \delta^\beta_\mu
 - g^{\beta \sigma } \delta^\nu_\alpha  \delta^\lambda_\mu
\nonumber \\
&&
 + g^{\beta \nu } \delta^\sigma_\alpha  \delta^\lambda_\mu
 + g^{\beta \sigma } g^{\lambda \nu } g_{\alpha \mu }
 - g^{\lambda \nu } \delta^\sigma_\alpha  \delta^\beta_\mu
 + g^{\nu \sigma } \delta^\lambda_\alpha  \delta^\beta_\mu
 - g^{\nu \beta } \delta^\sigma_\mu  \delta^\lambda_\alpha
 - g^{\nu \sigma } g^{\lambda \beta } g_{\alpha \mu }
\nonumber \\
&&
 + g^{\lambda \beta } \delta^\sigma_\mu  \delta^\nu_\alpha
 - g^{\nu \sigma } \delta^\sigma_\alpha  \delta^\beta_\mu
 + g^{\lambda \nu } g^{\sigma \beta } g_{\alpha \mu }
 + g^{\nu \beta } \delta^\sigma_\alpha \delta^\lambda_\mu
 - g^{\sigma \beta } \delta^\lambda_\mu \delta^\nu_\alpha
\biggr)
\end{eqnarray}

To get the diagonal form of the effective Lagrangian, we are to replace
the dynamical variables in the following way:

\begin{equation}
\gamma^a_{~ bc} \rightarrow  \bar{\gamma}^a_{~bc} = \gamma^a_{~bc} +
F^{-1a~~k}_{~~~~bc~~mn}
\biggl(  G^{~js~mn}_{p~~k} \nabla_s - T^{~j~mn}_{p~k} \biggr)
\lambda^p_{~j}
\label{repl}
\end{equation}
where $F^{-1 a~~k}_{~~~~bc~mn}$ satisfies three conditions

\begin{equation}
F^{-1 \sigma ~~\rho  }_{~~~~\mu \nu ~\alpha \beta } =
F^{-1 \rho  ~~\sigma }_{~~~~\alpha \beta  ~\mu \nu  }
\label{1}
\end{equation}

\begin{equation}
  F^{-1 \sigma ~~\rho  }_{~~~~\mu \nu ~\alpha \beta } =
- F^{-1 \sigma ~~\rho  }_{~~~~\nu \mu ~\alpha \beta } =
- F^{-1 \sigma ~~\rho  }_{~~~~\mu \nu ~\beta \alpha }
\label{2}
\end{equation}

\begin{equation}
F^{-1 \sigma ~~\rho  }_{~~~~\mu \nu ~\alpha \beta }
F_{ \rho~~ \omega  }^{~\alpha \beta ~\eta \kappa } =
\frac{1}{2} \delta^\sigma_\omega
\left( \delta_\mu^\eta \delta_\nu^\kappa
- \delta_\mu^\kappa \delta_\nu^\eta \right)
\label{3}
\end{equation}
The corresponding $F^{-1 \sigma ~~\rho  }_{~~~~\mu \nu ~\alpha \beta
}$, defined in paper ~\cite{MKL1}, have the following form:

\begin{eqnarray}
F^{-1 \alpha ~~\mu }_{~~~~\beta \sigma ~\nu \lambda }  & = &
\frac{1}{4} \Biggl(
  g^{\alpha \mu } g_{\beta \nu } g_{\sigma \lambda }
- g^{\alpha \mu } g_{\beta \sigma } g_{\nu \lambda }
+ g^{\alpha \mu } g_{\sigma \nu } g_{\beta \lambda }
- g_{\lambda \sigma } \delta^\mu_\beta \delta^\alpha_\nu
   + g_{\beta \sigma } \delta^\alpha_\nu \delta^\mu_\nu
\nonumber \\
&& + g_{\nu \lambda } \delta^\mu_\beta \delta^\alpha_\sigma
   - g_{\nu \beta } \delta^\mu_\lambda \delta^\alpha_\sigma
   + g_{\nu \beta } \delta^\mu_\sigma \delta^\alpha_\lambda
   - g_{\nu \sigma } \delta^\alpha_\lambda \delta^\mu_\beta
   - g_{\beta \lambda } \delta^\mu_\sigma \delta^\alpha_\nu
\Biggr)
\end{eqnarray}

The replacement (\ref{repl}) does not influence the functional
integral measure

\begin{equation}
det \left|
\frac{\partial \left( \lambda^a_{~\tau},
\bar{\gamma}^\sigma_{~\mu \nu } \right)}
{\partial \left( \lambda^b_{~\kappa}, \gamma ^\rho_{~\alpha \beta }
\right)} \right| = 1
\label{m}
\end{equation}

The effective Lagrangian (\ref{eff2}) is invariant under the general
coordinate transformation

\begin{eqnarray}
x^\mu & \rightarrow & 'x^\mu   =  x^\mu + {\it k} \xi^\mu(x)
\nonumber \\
e^a_{~\mu }(x) & \rightarrow & 'e^a_{~\mu }(x) =
- {\it k} \partial_\mu \xi^\nu e^a_{~\nu }(x)
- {\it k} \xi^\nu \partial_\nu e^a_{~\mu }(x) + O({\it k}^2)
\nonumber \\
\tilde w^a_{~b \mu }(x) & \rightarrow & 'w^a_{~b \mu }(x) =
- {\it k} \partial_\mu \xi^\nu \tilde w^a_{~b \nu }
- {\it k} \xi^\nu \partial_\nu \tilde w^a_{~b \mu }(x) + O({\it k}^2)
\label{coor}
\end{eqnarray}
and under the local Lorentz rotations

\begin{eqnarray}
x^\mu & \rightarrow & 'x^\mu   =  x^\mu
+ {\it k} \Theta^\mu_{~\nu }(x) x^\nu
\nonumber \\
e^a_{~\mu }(x) & \rightarrow & 'e^a_{~\mu }(x)
= {\it k} \Theta^a_{~b} e^b_{~\mu }(x) + O({\it k}^2)
\nonumber \\
\tilde w^a_{~b \mu }(x) & \rightarrow & 'w^a_{~b \mu }(x) =
{\it k} \Theta^a_{~c}  \tilde w^c_{~b \nu }(x)
- {\it k} \Theta^c_{~b}  \tilde w^a_{~c \mu }(x)
- {\it k} \partial_\mu \Theta^a_{~b} + O({\it k}^2)
\label{Lorentz}
\end{eqnarray}

The general coordinate invariance is violated by the following
gauge

\begin{eqnarray}
F_\mu & = & \frac{1}{2} \biggl( \nabla _\nu h^\nu_{~\mu}
+ \nabla_\nu h_\mu^{~\nu} -  \nabla_\mu h \biggr)
\nonumber \\
L_{gh} & = & \frac{1}{2} F_\mu F^\mu
\label{g1}
\end{eqnarray}

The action of the coordinate ghost is

\begin{equation}
L^{(coor)}_{gh}  = \overline{c}^\mu \left(
g_{\mu \nu } \nabla^\alpha \nabla_\alpha   + R_{\mu \nu } \right)
c^\nu
\label{gh}
\end{equation}

We fix the Lorentz invariance by means of the Landau gauge

\begin{equation}
f_{ab} = h_{ab} - h_{ba}
\label{g3}
\end{equation}

\begin{equation}
L_{gh}^{Lorentz}   = \lim_{\alpha  \rightarrow 0} \frac{1}{2 \alpha }
f_{ab} f^{ab} = \delta (f_{ab})
\label{g2}
\end{equation}
where $\delta(f_{ab} )$ is the delta-function.

The corresponding Lagrangian of the Lorentz ghost is

\begin{eqnarray}
L_{gh}^{(Lor)} & = &
\overline \omega_{ab} \frac{\delta f^{ab}}{\delta
\theta^{mn}} \omega^{mn}  \nonumber \\
& = & \overline \omega_{ab} \biggl(
\left( \delta^a_m e_n^{~b} - \delta^b_m e_n^{~a} - \delta^a_n e_m^{~b}+
\delta^b_m e_m^{~a} \right) \omega^{mn}  +
\nabla_a c_b - \nabla_b c_a \biggr) e
\label{gh3}
\end{eqnarray}
After some irrelevant redefinitions of the ghosts fields, we may drop
the term $\overline{\omega } c$ in equation (\ref{gh3}) as it alone is
insufficient for a closed-loop diagram containing
$\omega $ and $c$ fields.

Hence, the contribution of the Lorentz ghost to the one-loop effective
action is proportional to $\delta^4(0) $ and in the dimensional
regularization is equal to zero.

Summarizing all contributions we obtain that the one-loop counterterms
on the mass-shell including the contributions of the quantum and
the ghost fields are

\begin{equation}
\triangle ç_{\infty} = - \frac{1}{32 \pi^2 \varepsilon }
 \int d^4 x~ e \left(
\frac{19}{360}
R_{\mu \nu \sigma \lambda }(e) R^{\mu \nu \sigma \lambda }(e)
- \frac{89}{15} \Lambda^2 \right)
\label{newresult}
\end{equation}

This result does not coincide with the previous one (\ref{result}).

\section{Discussion}

Now we discuss the results of the previous chapters. Let us consider the
example $1$. At first, one needs to verify that the changes of the
variables (\ref{D2}) and (\ref{D3}) satisfy the condition (\ref{smat})
of the equivalence theorem.

Expressions (\ref{D2}) and (\ref{D3}) can be written in the
following form:

\begin{eqnarray}
\phi & = & \psi + O(\psi^2)
\nonumber \\
g_{\mu \nu}  & = & G_{\mu \nu} (1 + {\it k}^2 \psi^2 + O(\psi^4))
\label{DI3}
\end{eqnarray}

These changes of the variables satisfy the condition (\ref{smat}).
Hence, the equivalence theorem must be fulfilled. Then, the  question
arises:  what does the result (\ref{D8}) mean? We can suggest that such
situation when one field is quantized and not others
(the theory in the external field) is not consistent with the
equivalence theorem in nonrenormalizable theories.  Only when both
fields quantazing the equivalence theorem must fulfil.  Then, in
accordance with M.J.Duff \cite{Duff},  the result (\ref{D8}) can be
considered as a consequence of the inconsistency of quantum field
theory in an external gravitational field. Indeed, the actions
(\ref{D1}) and (\ref{D2}) describe the same classical theory written in
a different way.  Starting from the same classical theory written in
a different way, we obtain inequivalent quantum results on the
mass-shell (\ref{D8}).  There is no obvious principle which singles out
one particular choice of the classical action. An arbitrary classical
theory can be written in many different ways.  In the
semi-classical approach we cannot select one choice of variables as
"correct" and reject all the others.  Then the semi-classical approach
is inconsistent because it yields ambiguous results and one has no
criterion for deciding which is correct. In this way, inconsistency of
quantum field theory in the external gravitational field is the
consequence of the affirmation that the results of the loop
calculations on the mass-shell have some physical significance.
However, if we suggest that the results of the loop calculations within
the background field method on the mass-shell in the nonrenormalizable
theories are physically meaningless, then the result (\ref{D8}) has
a simple explanation. Both the actions (\ref{D1}) and  (\ref{D2}) are
equivalent at the classical and quantum domains.  But since the
one-loop counterterms on the mass-shell do not have physical
significance and do not give information about the S-matrix of the
theory, the demanding that $\triangle ç_{1 \infty}$ must be equal to
$\triangle ç_{2 \infty}$ is an additional, nonphysical request.  Both
the results (\ref{D5}) and (\ref{D6}) are true and both results are
physically meaningless.

Now we discuss example $2$. Since the results of the loop
calculations on the mass-shell depend on the gauge it is possible to
choose such a gauge that the theory described by the action (\ref{I1})
will be finite at the one-loop level. Hence, the affirmation that the
Einstein gravity interacting  with the matter fields (in particular,
scalar field)  is a nonrenormalizable (no-finite)  theory at the
one-loop level is wrong. This result can be explained by the
assumption that the results of the loop calculations on the mass-shell
do not have physical significance and do not give information
about the S-matrix of the theory. The problems connected with the
use of the gauge (\ref{I2}) and (\ref{I3})  have been
discussed also in paper \cite{MKL4}.
There is the other explanation of the result of example $2$
{}~\cite{odin}. The gauge (\ref{I3}) mixes the one-loop and the two-loop
order of perturbation theory. In order to obtain the gauge independent
result on the mass-shell in the gauge (\ref{I3}) one needs to take into
account the two-loop counterterms.

Recently the one-loop counterterms for Einstein gravity within the
class of gauge suggested in ~\cite{Ichi1} have been calculated in paper
{}~\cite{lavrov}. The resultant form for divergent part of one-loop
counterterms on the mass-shell does not coincide with the result of
paper ~\cite{Ichi1} and does not depend on gauge.

Let me discuss the results of the chapter $4$. Unlike example $1$ we
quantize both the fields existing in the theory, and the conditions of
the equivalence theorem are fulfilled. On the mass-shell the
term $\int d^4x~e~R_{\mu \nu \sigma \lambda }^2 $ can be rewritten as
$\int d^4x~e\left( R_{\mu \nu \sigma \lambda }^2  - 4 R^2_{\mu \nu }
+ R^2 \right) $. This expression is topologically invariant, so-called
Euler number, defined by

\begin{equation}
\chi   = \frac{1}{32 \pi ^2}
\int d^4x~e\left( R_{\mu \nu \sigma \lambda }^2  - 4 R^2_{\mu \nu }
+ R^2 \right)
\end{equation}
Then the results (\ref{result})  and (\ref{newresult}) can be written
in the following form

\begin{equation}
\triangle ç_{1 \infty}  = - \frac{1}{\varepsilon } \left(
\frac{53}{45} \chi  + \frac{29 {\it k}^2 \Lambda S}{160 \pi^2}
\right)
\label{81}
\end{equation}

\begin{equation}
\triangle ç_{1 \infty}  = - \frac{1}{\varepsilon } \left(
\frac{19}{360} \chi  + \frac{89 {\it k}^2 \Lambda S}{960 \pi^2}
\right)
\label{82}
\end{equation}
where $S$ is the classical action on the mass-shell.

In the topological trivial space-time $(\chi = 0)$, the considered
theory described by the action (\ref{action}) or (\ref{action2}) is
renormalizable on the mass-shell. Two different
sets of dynamical variables give rise to different renormalization
group functions

\begin{eqnarray}
\mu^2 \frac{\partial \bar \lambda}{\partial \mu^2}  & = &
- \frac{29}{160} \bar \lambda^2
\label{eq1}  \\
\mu^2 \frac{\partial \bar \lambda}{\partial \mu^2}  & = &
- \frac{89}{960} \bar \lambda^2
\label{eq2}
\end{eqnarray}
where $\lambda = {\it k}^2 \Lambda$ is the dimensionless constant and
equations (\ref{eq1}) and (\ref{eq2}) are connected with the results
(\ref{81}) and (\ref{82}), respectively.

It has been argued in paper ~\cite{ChD1} that in the topological
non-trivial space-time $(\chi \neq 0)$ to obtain the one-loop
renormalizable theory the term $\int d^4x~e\left( R_{\mu \nu \sigma
\lambda }^2  - 4 R^2_{\mu \nu } + R^2 \right) $ must be added  to the
classical action with the coefficient $\alpha $. Since $\chi $ is
topologically invariant, this can be done without damage to the field
equation and one-loop counterterms in the space-time without
boundaries. Then, the one-loop counterterms can be absorbed into a
renormalization of the new topological coupling constant $\alpha $ and
cosmological constant $\Lambda $. The renormalization group equations
describing the behaviour of the topological constant are the following:

\begin{eqnarray}
\mu^2 \frac{\partial \alpha }{\partial \mu^2}  & = &
- \frac{106}{45}
\label{new1}  \\
\mu^2 \frac{\partial \alpha }{\partial \mu^2}  & = &
- \frac{19}{180}
\label{new2}
\end{eqnarray}
where equations (\ref{new1}) and (\ref{new2}) are connected with the
results (\ref{81}) and (\ref{82}), respectively.

Two sets of the dynamical variables
$(e^a_{~\mu}, Q^\sigma_{\mu \nu})$ and $(e^a_{~\mu}, \tilde
\omega^a_{~b\mu})$ must be equivalent at the classical and quantum level
because the transformation from one to an other set of the
variables does not change the functional integral measure

\begin{equation}
det \left| \frac{\partial (e^a_{~\mu}, \tilde \omega^a_{~b \mu})}
{\partial (e^m_{~\sigma}, Q^\lambda_{~\alpha\beta})} \right| = 1
\end{equation}

Instead of the dynamical variables $(e^a_{~\mu }, Q^a_{~\mu \nu })$ or
$(e^a_{~\mu }, \tilde \omega^a_{~b \mu } )$ we can consider the other
two sets $(g_{\mu \nu }, Q^\sigma_{~\mu \nu })$ or
$(g_{\mu \nu }, \tilde \Gamma^\sigma_{~\mu \nu } )$ where the
connection $\tilde \Gamma^\sigma_{~\mu \nu } $ satisfies the metric
condition (\ref{metric}). The results of the one-loop calculation
within these variables will coincide with (\ref{result}) and
(\ref{newresult}) respectively.

The one-loop counterterms in the first-order gravity with the
Gilbert-Einstein action without the cosmological constant have been
calculated in the paper~\cite{shapiro}. However in our opinion, in this
paper there is inconsistency between the equations of motion and the
choice of the dynamical variables. In the affine-metric theory with the
metric and connection fields as independent dynamical variables the
tensor of connection defect ($A^\sigma_{~\mu \nu }$ in the notations of
paper ~\cite{shapiro}) is not equal to zero (see ~\cite{MKL2},
{}~\cite{MKL3}).

The classical Lagrangian (\ref{action2m}) written in the two different
classically equivalent ways give rise to different quantum results.  We
can introduce in the theory, described by the action (\ref{action}) or
(\ref{action2}), the matter fields interacting only with the vierbein
(or metric) field. In this theory, the torsion fields will be
auxiliary, nonpropagating fields. However, the results of the loop
calculations will also depend on the choice of dynamical variables.

\section{Conclusion}

The main aim of this paper was to show that the loop calculations in
the nonrenormalizable quantum gravity are ambiguous ones.
It has been investigated in the previous chapters that the
classical theory written in a different way leads to the inequivalent
quantum results depending on the choice of dynamical variables,
gauge fixing term and the choice of parametrization. The modern
point of view is that the physical observation quantities must be
independent of the  choice of such non-physical parameters as gauge and
parametrization. As a consequence, we obtain that the results of the
loop calculations depending on the nonphysical parameters must be
physically meaningless. Due to violation of the equivalence
theorem and DeWitt-Kallosh theorem in the nonrenormalizable theories of
the quantum gravity, the question arises about the criteria of
a sensible theory of the quantum gravity. It is possible that an
arbitrary theory nonrenormalizable by power counting like the
Einstein gravity can be finite order by order in  perturbation
theory by the choice of the non-standard gauge fixing term.

The other question is what will be with the renormalizable theory of
the quantum gravity with higher derivatives ?  The calculations
like example $2$ with the non-standard gauge do not take the place. All
one-loop counterterms in this theory have been calculated only in
the Landau-DeWitt gauge~\cite{Renor}. The validity of the
DeWitt-Kallosh theorem in the theory of the quantum gravity with
higher derivatives must be verified by the calculations of the one-loop
counterterms in the arbitrary parametrization and by means of the gauge
distinct from the Landau-DeWitt gauge ~\cite{MKL4}. However, the
example $1$ and chapter $4$ take the place also in the renormalizable
theory of the gravity. It has been shown ~\cite{scalar1},
{}~\cite{scalar2} that the necessity to introduce the term $\xi \phi^2 R$
is demanded by the renormalizability of the gravity interacting with
the scalar fields. Hence, in the renormalizable theory of gravity
the ambiguity of the loop calculations will also be present.

What is the reason for these strange results ? All methods and theorems
of quantum field theory are based on some principles like
renormalizability and unitary. All existing theories of gravity
(based on the Riemannian and non-Riemannian space-time structure with
and without supersymmetry) do not satisfy these principles. There
are unitary, but nonrenormalizable theories (like the Einstein gravity)
or renormalizable, but non-unitary theories (like the theory with
higher derivatives) or nonrenormalizable and non-unitary theories.
Hence, all existing theories of quantum gravity are unsatisfactory
>from the point of view of quantum field theory . We suggest that in an
arbitrary existing theory of the quantum gravity all results of the
loop calculations do not have physical significance. In our opinion,
to construct a sensible theory of the quantum gravity, one
needs to use non-standard methods of calculation, for example,
non-perturbative methods of  calculation of the quantum
corrections.

The only way to avoid this ambiguity is to suggest
that loop counterterms on the mass-shell in the nonrenormalizable
theories are physically meaningless. Then, the results (\ref{result})
and (\ref{newresult}) have the same physical ground.

To summarize, the only way to explain the results (\ref{D8}),
(\ref{I4}) and (\ref{newresult}) is to take that in the
nonrenormalizable theories of the gravity the results of the loop
calculations on and off mass-shell do not have physical
significance. As a consequence, all physical predictions and
calculations performed on the basis of the loop calculations in the
nonrenormalizable theories of the quantum gravity are meaningless.

We are very grateful to L. V. Avdeev, D. I. Kazakov, D.Fursaev,
S. Solodukhin (JINR, Dubna) and colleagues from Moscow State University
for many useful discussions. We are greatly indebted to G.Sandukovskaya
for critical reading of the manuscript.

\end{document}